\documentstyle[psfig]{aa}

\input epsf         
\begin{document}

\title{Size-Mass-luminosity relations in AGN\\ and the role of the
accretion disc}

\titlerunning{The Size-Mass-Luminosity Relations in AGN.}

\author{Suzy Collin\inst{1} and Jean-Marc Hur\'e\inst{1,2}}

\institute{$^1$DAEC/UMR 8631 du CNRS, Observatoire de Paris, Section de
Meudon, F-92195 Meudon\\
$^2$Universit\'e Paris 7 Denis Diderot, Place Jussieu, F-75251 Paris
Cedex 05}

\thesaurus{02.01.2, 02.09.1, 11.01.2, 11.14.1, 11.17.3,
12.05.1,19.37.1,19.92.1}

\offprints{Suzy Collin (suzy.collin@obspm.fr)}

\maketitle

\begin{abstract}
We address the question of the relations between the black hole's mass,
the accretion rate, the bolometric
luminosity, the optical luminosity and the size of the Broad Line
Region (BLR) in Active Galactic Nuclei, using recent observational
data obtained from monitoring campaigns. We show that a
standard accretion disc cannot account for the observed optical luminosity,
  unless it radiates at
super-Eddington
rates. This implies the existence of another, dominant emission mechanism
in the optical range, or a non standard disc (non stationary, ADAF and/or
strong outflows).
  Narrow Line Seyfert 1 galaxies (NLS1s) are most extreme in this context:
  they have larger
bolometric to Eddington luminosity ratios than Broad Line Seyfert 1
(BLS1s), and most likely a larger ``non disc" component in the optical
range. From realistic
simulations of self-gravitating $\alpha$-discs, we have systematically
localized the gravitationally unstable disc and shown that, given
uncertainties on both the model and observations, it coincides quite well with
  the size of the BLR. We
therefore suggest that the gravitationally unstable disc is the source
which releases BLR clouds in the medium. However the influence of the
ionization parameter is also required to explain the correlation found
between the size of the BLR and the luminosity. In this picture the size of the
  BLR in NLS1s (relative to the black hole size) is
larger (and the emission line width smaller) than in BLS1s simply because
their Eddington ratio is larger.


\end{abstract}

\section {Introduction}

There are several ubiquitous media in the central region of Active
Galactic Nuclei (AGN): surrounding the supermassive black hole (BH), there
are
an accretion disc (AD), an X-ray source, a photoionized Broad Line Region
(BLR)
and possibly a molecular torus at the scale of a few parsecs from the
center (Antonucci \& Miller 1985). Despite years of efforts, the physics
of these
components is far from being completely understood as well
as the possible interaction between them. For instance, the origin of the
observed correlation between the size of the BLR and the luminosity of the
AGN, which, under some hypothesis, translates into a relation between the
mass of the central black hole and the disc accretion rate, is still
mysterious.

In this paper, we try to understand the links between the accretion disc
(at small and large radii), the optical, bolometric and Eddington
luminosities, and finally the Broad Line Region (BLR). For this purpose, we
use recent observational data which have led to the determination of the
size of the BLR, optical luminosity, and black hole (BH) mass in a few tens of
objects spanning a large range of luminosity (Kaspi et al. 2000). In
Section 2, we briefly recall the emission mechanisms in Active Galactic
Nuclei (AGN), their link with the accretion disc, the main properties of
the
disc itself and those of the BLR. In Section 3 we discuss the
relations between the observed optical luminosity, bolometric
luminosity and BH mass. It is demonstrated that the standard disc
emission cannot
   account for the optical luminosity, and that Narrow Line Seyfert 1
galaxies (NLS1s) are the most extreme cases on this point of view. In Section
4, we model the outer disc and show that the occurrence of gravitational
instability roughly coincides with the size of the BLR. Our conclusions are
presented in the last section.

\begin{figure*}
\qquad \psfig{figure=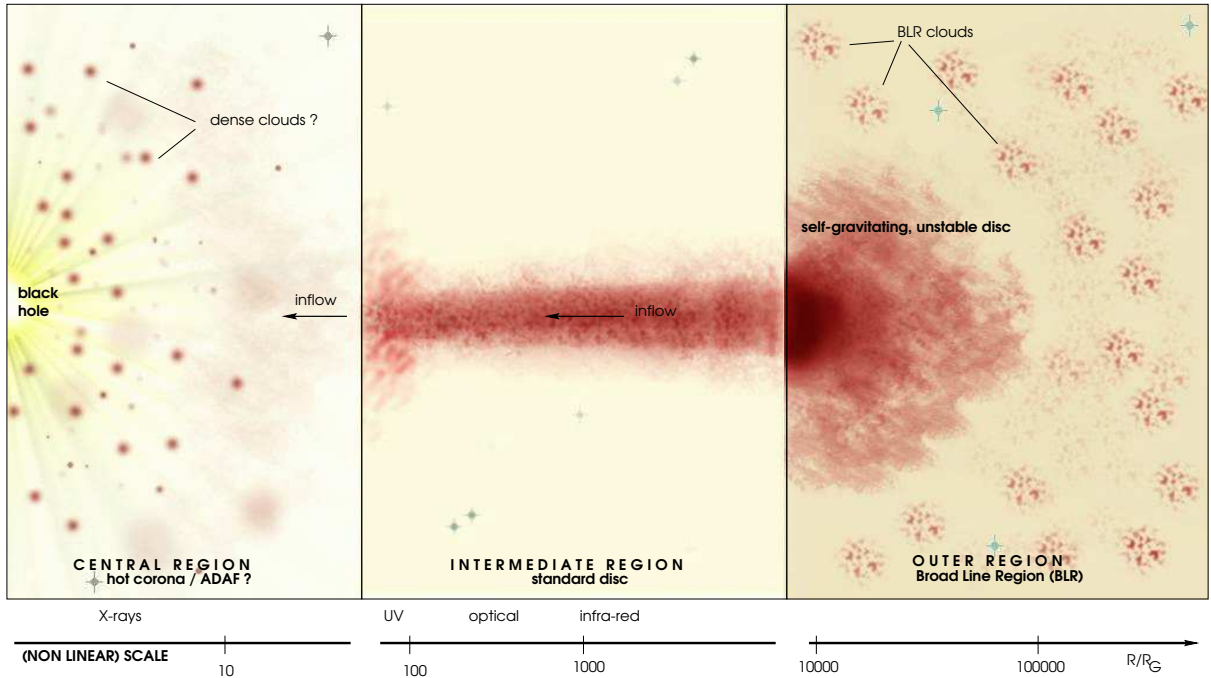,width=16cm,angle=0}
\caption{Schematic view of the most central region of an AGN. The scaling
depends slightly on the mass and on the accretion rate, and is appropriate
for a
$10^8$ M$_\odot$ black hole ($R_{\rm G} \simeq 1.5 \times 10^{13}$ cm)
  accreting at
$\dot{m}\sim 0.1$ in Eddington units.}
\label{fig-disque}
\end{figure*}

\section{General considerations}

\subsection{Emission mechanisms in AGN: the inner disc}

Since its discovery in AGN spectra, the universal feature
known as the ``Big Blue Bump" (BBB) which represents the bulk of the
bolometric luminosity, has been
interpreted as thermal emission of an accretion disc (Shields
1978, Malkan \& Sargent 1982). If this disc is Keplerian and optically
thick as generally assumed (the so-called ``standard disc" model), then
its emission reproduces roughly the BBB feature (see for a review 
Koratkar \& Blaes 1999,
  and Collin 2000). Moreover, the idea that a non thermal power law 
continuum dominates
the near infrared band and contributes, together with the disc, to the
total emission in the optical range, has been abandoned after the
publication of the generic spectrum of radio quiet
quasars by Sanders et al. (1989). These authors have actually shown that
the IR band is most likely
due to hot dust emission peaking at a few microns. Since hot
dust
sublimates at temperatures $\sim 1700$ K, dust cannot
contribute to the optical emission.
So, {\it if one sustains the standard disc picture}, one must admit that
the optical continuum component is entirely emitted by the accretion disc.

According to the standard model, the disc regions emitting in the optical
band are located at a few 100 $R_{\rm G}$, where $R_{\rm G}=GM/c^2$ is
the gravitational radius of the BH with mass $M$ (see Fig.
\ref{fig-disque}, {\it middle panel}). Since the accretion luminosity
varies as $R^{-1}$, the optical luminosity should be only a small fraction
($\sim 1\%$) of the bolometric luminosity.  In contrast, the
regions emitting the UV and EUV components are located close to the BH,
typically at distances of a few 10 $R_{\rm G}$. Modeling the disc
spectrum in the UV and EUV ranges is
a difficult task as radiation is subject to several processes
such as Comptonization. The emergent spectrum
does depend strongly on the details of the disc vertical stratification
which is not well known.
On the contrary, in the
optical range, one can show that the local spectrum is roughly
  a blackbody at the effective temperature corresponding to the
conversion of gravitational energy into radiation (Collin 2000).

The bolometric luminosity $L_{\rm bol}$ is a robust prediction for a
(steady state) standard disc: it is related only to the accretion rate
$\dot{M}$ and to the radius of the last, dynamically stable orbit, and
corresponds to an efficiency of mass-energy conversion $\eta \approx 10\%$
(the value adopted throughout this paper), depending on the spin of the
black hole. Thus:
\begin{equation}
L_{\rm bol}=\eta \dot{M} c^2,
\label{eq:lboldef}
\end{equation}
and in principle, accretion stops if $L_{\rm bol} > L_{\rm Edd}$
where $L_{\rm Edd} \simeq 1.5 \times 10^{45} M_7$ erg/s is the Eddington
luminosity and $M_7$ is the mass in units of $10^7$ M$_\odot$.

AGN spectra also display a hard X-ray spectrum which extends up to a few
hundred keV. This hard component which contributes a significant fraction of
the bolometric luminosity (typically 30$\%$) cannot be produced by a bare
standard disc. This is why one generally invokes the presence of a hot gas
close to the BH (see Fig. \ref{fig-disque}, {\it left panel}), either in
the form of a corona above the disc, or, instead of the inner disc itself,
a quasi-spherical medium. In this case, the luminosities emitted
respectively in the
X-ray and optical-UV ranges relative to the accretion luminosity depend on
the radius where this hot medium physically connects to the outer,
standard disc.

The likely change in the morphology of the accretion flow in the inner
region (thin disc $\rightarrow$ thick disc/spherical medium) is predicted
 by theory. The standard model probably does not work at small
radii where the disc is radiation pressure supported because of thermal
instabilities. In comparison, the standard disc as a model
for outer regions ($R \sim$ a few $10^2 R_G$) seems more consistent:
the disc is sustained by gas pressure, relatively cold, and much more
stable. It is true that a thermal instability, possibly recurrent
(Siemiginovska, Czerny \& Kostyunin 1986), is expected where hydrogen
recombines/ionizes but such an instability is predicted to keep the disc
in a geometrically thin configuration.

\subsection{The status of the outer, self-gravitating disc}

At large radii (roughly $R \ge 10^3 R_{\rm G}$; see below), the
standard disc solution must be modified
   to account for the disc vertical self-gravity which may exceed the
vertical component of
the central gravity. The occurrence of vertical self-gravity is
inescapable since it varies roughly as $R^{-3/5}$, that is less steeply
than central gravity which varies as $R^{-3}$ (Shlosman \& Begelman
1998; Hur\'e et al. 1994; Hur\'e 1998). When the hydrostatic equilibrium
is no longer controlled by the central object but by the disc, the surface
density continuously increases whilst the disc gets thinner and
thinner. But a self-gravitating disc is expected to be gravitationally
unstable as soon as some criterion is fulfilled (Goldreich \&
Lyndell-Bell 1965; see also Toomre 1964), usually this is $Q =
{\Omega^{2}\over \pi G \rho} \le 1$, where $\Omega$ is the Keplerian
angular velocity and $\rho$ is the local mass density.
 
The structure of the disc in the gravitationally unstable region is not
known. It may become marginally stable again, if the viscosity
self-adjusts in some fashion (Lynden-Bell \& Pringle 1974).
Gravitational instabilities increase random (radial and vertical)
velocities and could
lead to the disc fragmentation into clouds (Pacynski 1968; Shore \&
White 1982, Shlosman \& Begelman 1989). Following this scenario (see Fig.
\ref{fig-disque}, {\it right panel}), it is often
assumed that these clouds are
    moving with a high velocity, and that their chaotic motion and
    mutual collisions provide the support against vertical gravity and keep
accretion at the required rate (Begelman \& Krolik 1986, Kumar 1999). The
clumpy disc could become geometrically thick again, and even, at very
large distance ($\sim 1$ pc),
it could connect to the dusty molecular torus invoked in the Unified
Scheme of
Seyfert nuclei (Antonnucci \& Miller 1985). It is also
possible that fragments in the unstable disc collapse to form protostars.
Because these protostars can trap a large amount of gas from the disc, they
should rapidly evolve into massive stars, and then give rise to outflowing
gas through winds and supernovae explosions (Collin \& Zahn 1999). In
both scenarios one would thus expect the presence of a dense medium above
the accretion disc,
with an azimuthal velocity close to the Keplerian one, which could be
identified with the BLR.

\subsection{Size and velocities within the BLR}

A major tool for studying the structure of the BLR is
reverberation mapping, through the study of
correlated variations of the lines and continuum fluxes
    (see the review by Peterson 1993). It has been used to
determine the size of the BLR in several tens of AGN, and to correlate the
mass of the central black hole with the luminosity of the AGN, under the
assumption that the BLR is gravitationally bound to the black hole.

Though powerful for size determination, reverberation
mapping does not provide the velocity field of the BLR, and
can generally not distinguish between
  radial and rotational motions. In some
objects however, there are indications that rotational motions of the
medium emitting the H$\beta$ line dominates over radial motions. On the
other hand, the
medium emitting high ionization lines (or HILs) is sometimes observed as
an outflowing medium. This is thought to be the case in NGC 4051 (Peterson et al. 2000). However 
it is likely
that the velocity $V_{\rm BLR}$ does not differ strongly
from a gravitationally bound, Keplerian or virial motion (Peterson \&
Wandel
1999). Then, the radius of the BLR is
approximately
given by
\begin{equation}
{R_{\rm BLR} \over R_{\rm G}} = \frac{GM}{V^2_{\rm BLR}} \times
\frac{c^2}{GM} \sim {900 \over V_9^2}
\label{eq-RBLR}
\end{equation}
where $V_{\rm BLR} = V_9 \times 10^9$ cm/s.

Because there is no consensus about the dynamics of the BLR, there has been
little progress in understanding the physics of this region. The most
commonly accepted picture, deduced from photoionization models, consists
of an assembly of photoionized clouds that cover at least $10\%$ of the
source emitting the incident, primary continuum. The number of BLR clouds
is
probably very large because the volumic filling factor of the emitting
medium is extremely small. This large population ensures a large
total emitting area as well as a high density and a small
individual size. The smoothness of the line profiles also
requires a large number of clouds. If
these clouds are not confined by any mechanism, they expand in a very
short time scale (less than one year). If the BLR is stationary, new
clouds must be permanently generated. Alternatively, if the the clouds are formed
uniquely then some form of confinement mechanism must operate. The thermal confinement by a hot
medium as proposed in the "two phase model" by Krolik, Tarter \& Mc Kee
(1981)
implies a Thomson thick hot medium, which is difficult to reconcile with the very short
  variation time scale of the
X-ray flux.

A natural idea which does not address the confinement problem is that the
BLR is made of the atmosphere of giant or ``bloated stars'', or by winds
from giant stars (Edwards 1981,
  Penston 1988, Scoville \& Norman 1988, Alexander \&
Netzer 1994). This model however requires a very large number of
stars to account both for the line luminosity and for the smoothness of
the line profiles. Another possibility that does not appeal to any special
confinement mechanism either is that the lines are emitted by the accretion
disc.  Dumont \& Collin-Souffrin  (1990) have proposed that low 
ionization lines
  (or LILs), like
Balmer lines, which require a high density emitting medium
(Collin-Souffrin et al., 1986) are formed at the surface of the accretion
disc. On the other hand,  Murray \& Chiang (1995,
1997) proposed that the broad emission lines are emitted by a wind
released at the top of the disc where motions are still mainly rotational.
In this model, the broad absorption lines (BALs) are produced by the 
high-velocity component
of the wind. In both cases, the BLR is dominated by rotation and its
presence (or at least, that of the region emitting the LILs) is linked
to the accretion disc.

\section{Relation between the optical luminosity and the bolometric
luminosity}

\subsection{Global investigation}

Kaspi et al. (2000) combined their own reverberation mapping study
with other published data to determine the size of the BLR, the BH   mass
and the optical luminosity. As it is usually done (Peterson 1993, 
2000), they derived
  the size of the BLR from the
measurements of the time lags between the light curves relative to the
H$\alpha$ and H$\beta$ fluxes and the light curve of the underlying
continuum which is assumed
to follow the ionizing continuum. From the size of the BLR, they computed
the BH mass
  from the FWHM of H$\beta$ and H$\alpha$, assuming that these lines
are emitted by a gravitationally bound medium. There is some uncertainty
in the mass determination (by a factor of a few) since there are two
different, non-equivalent methods to determine the FWHM of lines.
The FWHM can be averaged on all spectra, giving a ``mean'' BH
mass: this is the method preferred by Kaspi
et al. (2000). In the method proposed by Peterson et al (1998), the FWHM is
measured on the rms spectrum, giving a ``rms'' mass.

\begin{figure}
\begin{center}
\psfig{figure=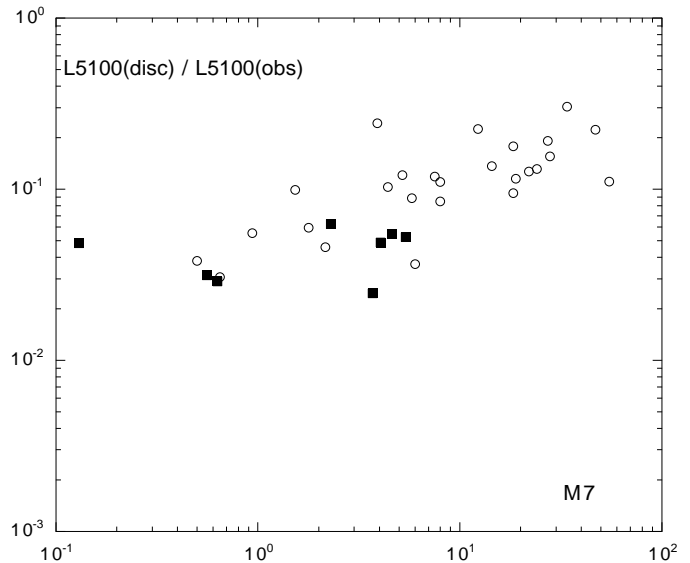,width=9cm}
\caption{$L_\nu^{\rm disc}(5100\AA)/L_\nu^{\rm obs}(5100\AA)$ ratio 
versus the mass (in units of
10$^7$M$_{\odot}$)
for the Kaspi et al. (2000) sample:  $L_\nu^{\rm disc}(5100\AA)$ have been
computed using Eqs. (\ref{eq:lboldef}) and (\ref{eq:9}). We have distinguished
NLS1 galaxies ({\it squares}) and other nuclei ({\it circles}).}
\label{fig-L5100calcobs-vs-M}
\end{center}
\end{figure}

\begin{figure}
\begin{center}
\psfig{figure=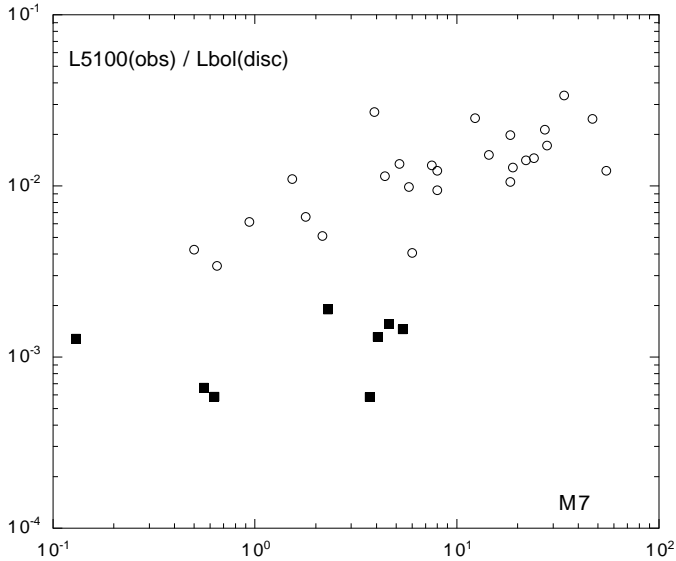,width=9cm}
\caption{$L(5100\AA)_{\rm obs}/L_{\rm bol}$ ratio versus the mass (in units of
10$^7$M$_{\odot}$)
for the Kaspi et al. (2000) sample: $L_{\rm bol}$ has been computed 
assuming that
the optical luminosity $L(5100\AA)_{\rm obs}$
is entirely due to the disc. We have distinguished
NLS1 galaxies ({\it squares}) and other nuclei ({\it circles}).}
\label{fig-L5100Lbolcalc-vs-M}
\end{center}
\end{figure}

\begin{figure}
\psfig{figure=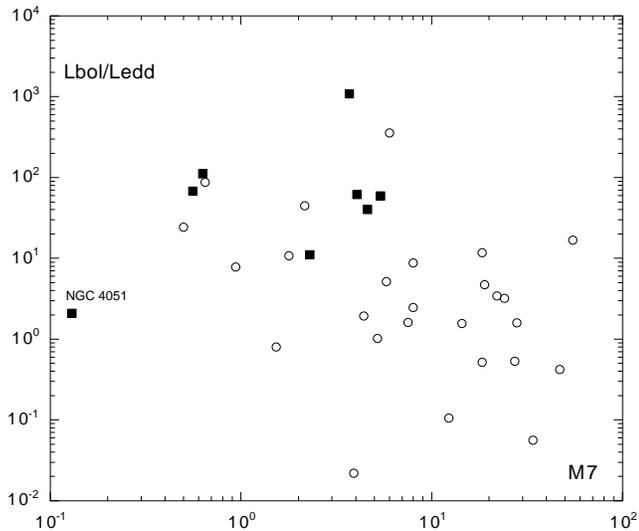,width=8.5cm,angle=0}
\caption{$L_{\rm bol}/L_{\rm Edd}$ ratio versus the mass (in units of
10$^7 $M$_\odot$), for the Kaspi et al. (2000) sample, assuming that the
optical
monochromatic luminosity $L^{\rm obs}(5100\AA)$ is entirely due to the
disc.  We have distinguished
NLS1 galaxies ({\it squares}) and other nuclei ({\it circles}). Note the point
corresponding to the lowest mass is NGC
4051.}
\label{fig-LbolcalcLEdd-vs-M}
\end{figure}

To estimate the bolometric luminosity from the monochromatic luminosity,
$L_{\rm bol} \propto \nu L_\nu^{\rm obs}(5100\AA)$ is generally assumed.
For the AGN generic spectrum (Laor et al. 1997), one deduces:
\begin{equation}
L_{\rm bol} \sim 9 \times \nu L_\nu^{\rm obs}(5100\AA).
\label{eq:9}
\end{equation}
  For each object
in Kaspi et al. sample, we have compared
  this optical-to-bolometric conversion with two {\it non equivalent but both
self-consistent} methods of conversion based on the disc model:
\begin{enumerate}
\item by computing the theoretical, monochromatic luminosity $\nu
L_\nu^{\rm disc}(5100\AA)$ from the spectrum of an optically thick, steady
state disc given the pair $(M,\dot{M})$, assuming Eq.(\ref{eq:9}), and
Eq.(\ref{eq:lboldef}) to estimate $\dot{M}$.
Figure \ref{fig-L5100calcobs-vs-M} displays the ratio $L_\nu^{\rm
disc}(5100\AA)/L_\nu^{\rm obs}(5100\AA)$  obtained with this method. 
It shows that,
for all objects
in the sample, the emission due to the disc can represent only a minor, and in
most cases, a negligible fraction of the observed luminosity in the
optical band.
\item by finding the ``right'' accretion rate $\dot{M}$ (and then $L_{\rm
bol}$ via Eq.(\ref{eq:lboldef})) such that the observed monochromatic
luminosity is only due to a standard disc (that is $L_\nu^{\rm
disc}(5100\AA) = L_\nu^{\rm obs}(5100\AA)$).
  This method leads to extremely small ratios $\nu L_\nu^{\rm
obs}(5100\AA)/L_{\rm bol}$, as shown by Fig. \ref{fig-L5100Lbolcalc-vs-M}. It implies a very large bolometric 
luminosity,
and therefore a very large Eddington ratio $\dot{m} = L_{\rm bol}/L_{\rm Edd}$.
  To illustrate this last point the
corresponding
Eddington ratio is plotted versus the
central mass on Fig. \ref{fig-LbolcalcLEdd-vs-M}. It shows that almost
all the objects would be {\it radiating at highly super-Eddington rates},
proving that this second self-consistent optical-to-bolometric 
conversion is wrong.
\end{enumerate}

  In other words, {\it a relation of the form
Eq.(\ref{eq:9}) should be roughly valid}, meaning that {\it the optical
luminosity  is not radiated by a standard disc}.

Note that we assume a face-on disc, thus overestimating slightly $L_\nu^{\rm
disc}$. These conclusions are unchanged if the inner disc is removed for
some reason, that is, if the inner radius is larger (in the limit of a few
10 $R_{\rm G}$) than a few $R_{\rm G}$ as considered here, since the
optical emission is almost entirely produced in outer regions located much
further away from the center. Clearly the situation becomes even more 
extreme if
the
region emitting the optical band is not present in the disc, that is, if
the disc is truncated for some reason, for instance due to self-gravity.

There are at least four possible explanations for this result:

\begin{enumerate}
\item the optical luminosity is not due to the disc, as quoted. This is
plausible since the standard disc spectrum is well known to be flatter
than the spectrum of Seyfert 1 nuclei in the optical range
  (Koratkar \& Blaes 1999, Hubeny et
al. 2000, Collin 2000). This is also compatible with a recent study of the
NLS1 1 RE J1034+396 by Puchnarewicz et al. (2000) who conclude that an
additional underlying power law connecting the IR to the X-rays is really
necessary to fit the optical/UV continuum. They suggest that this extra
component might be a non thermal BLac-type component. This is indeed
possible if for instance
there is a strong magnetic
   field anchored into the disc. But it could also be due to a thermal
process like the emission of dense clouds surrounding the X-ray source
and reprocessing a large fraction of the X-ray luminosity, as proposed by
Celotti, Fabian and Rees (1992), and by Collin-Souffrin et  al. (1996).
This cloud system should be distributed quasi-spherically in order to
reprocess a large fraction of the X-ray luminosity (in the case of a disc,
the reprocessed fraction decreases as $R^{-1}$, like the accretion
luminosity, and is necessarily small in the optical range).

\item the standard disc model is not valid. Although the accretion
luminosity does not depend on viscosity, the fact that the disc is
effectively optically thick and radiates like a black body is not
independent of the viscosity prescription. Note, as a proof, that the
optical thickness of a standard disc is a function of the
$\alpha$-viscosity parameter. Either a completely different mechanism of
angular momentum evacuation in the disc is at work (e.g. density waves),
or the formalism for turbulent viscosity must be reconsidered, or both.

\item the accretion rate is not uniform. This possibility includes the
presence of a non steady disc and the possible existence of some steady
inflow involving an accretion rate decreasing with decreasing radius.
Concerning the first point, there are a few well known sources of
instability that can perturb the equilibrium of a disc but this would mean
that all objects are observed in a ``high optical state". This is
statistically somewhat doubtful. Regarding the second point, this would
mean that a huge amount of gas is accreted at large distances and then
converted into an outflow at smaller radii. This is also unrealistic
because it would imply massive super-Eddington outflows, which are not
observed.

\item accretion is advection dominated in the inner regions. Such a
solution is still not demonstrated on theoretical grounds. Stable
Advection Dominated Accretion Flows (ADAFs) (Narayan \& Yi 1995, and
many other works) take place for relatively low accretion rates, well
below the Eddington limit.  At the other extreme, there are advective
solutions for Eddington rates (``slim discs" for instance, cf. Abramowicz
  et al. 1988) or for super-Eddington rates (thick discs), but it
is difficult to conceive a
durable, super-Eddington accretion rate, as it would imply a dramatic
growth of the hole over cosmic time scales. In conclusion, even if the disc
  becomes
geometrically thick in the inner regions, advection is not likely to be
important in the Seyfert galaxies and moderately luminous quasars
considered here, and  the gas should flow all the way down to the last
stable orbit and radiate accordingly.
\end{enumerate}

\subsection{The case of Narrow Line Seyfert 1}

\begin{figure}
\psfig{figure=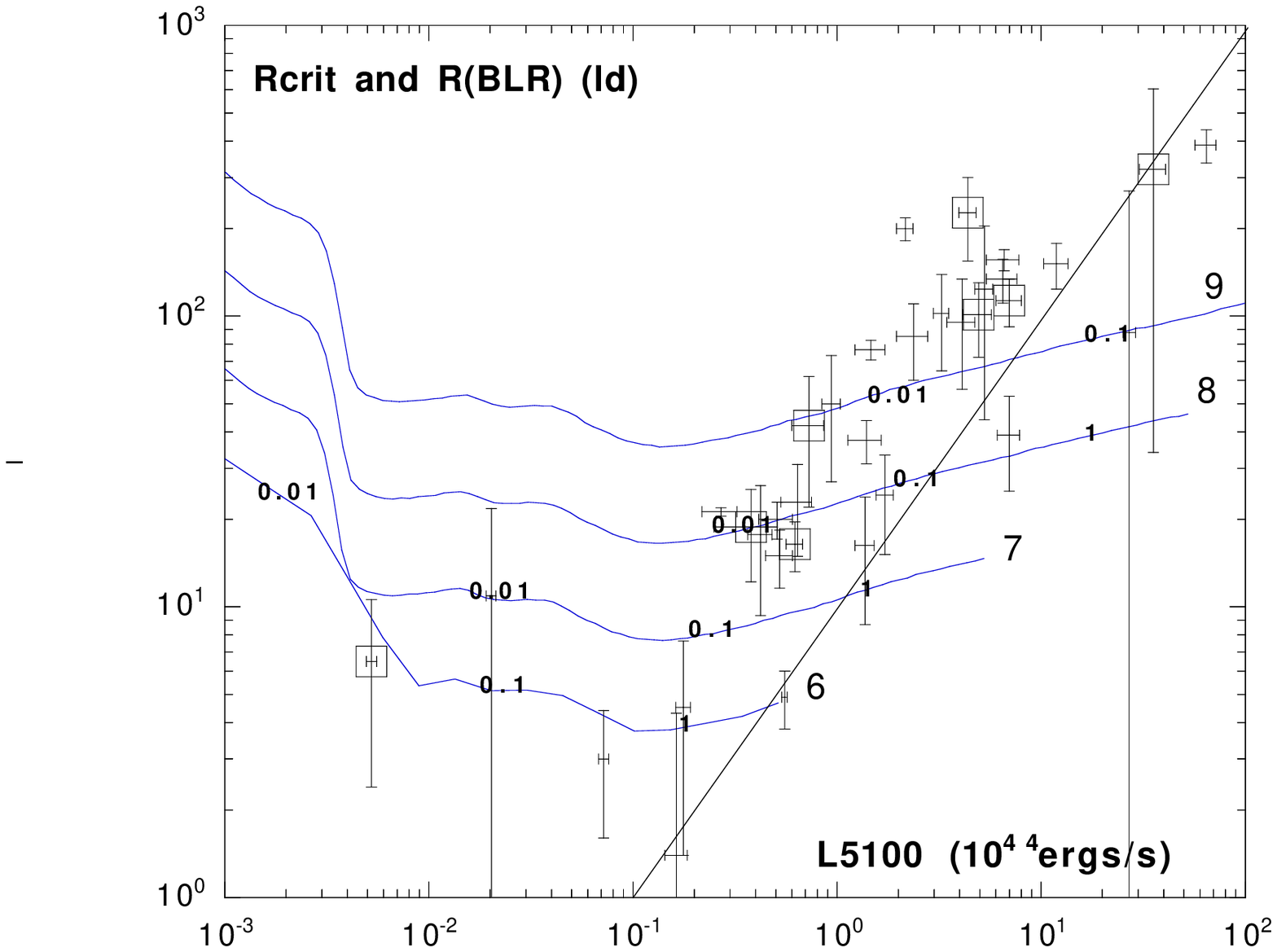,width=8.5cm,angle=0}
\caption{The points give $R_{\rm
BLR}$ in lt-days versus the monochromatic luminosity at 5100 $\AA$ in
units of 10$^{44}$ erg s$^{-1}$, for the Kaspi et al. (2000) sample. The big open
squares
show NLS1s.
The curves give $R_{\rm
crit}$ from the 2D
simulations. The accretion rates of the theoretical curves have been
converted into
$L$5100 assuming that the bolometric luminosity  is
equal to 9$\times L$5100. The curves are labeled with the black hole
mass in log(M$_\odot$), and the
Eddington ratio
$\dot{m}$ is indicated on the curves. }
\label{fig-R-vs-L5100}
\end{figure}

Figure \ref{fig-R-vs-L5100} displays the size of the BLR, $R_{\rm BLR}$,
  versus $\nu L_\nu^{\rm
disc}(5100\AA)$ for the Kaspi et al. (2000) sample. In this figure and
in Figs.  \ref{fig-L5100calcobs-vs-M}, \ref{fig-L5100Lbolcalc-vs-M}, and
\ref{fig-LbolcalcLEdd-vs-M},
NLS1 have been isolated\footnote{We define NLS1 as Seyfert nuclei with
$V_{\rm BLR} = (GM/R_{\rm BLR})^{1/2} < 2000$ km/s, instead of using
   FWHMs, to avoid having
to take into account a correction for
H$\alpha$ and H$\beta$}. We notice that they do not occupy any privileged
region on Fig. \ref{fig-R-vs-L5100}, contrary to Figs.
  \ref{fig-L5100calcobs-vs-M}, \ref{fig-L5100Lbolcalc-vs-M} and
\ref{fig-LbolcalcLEdd-vs-M}. On
Fig. \ref{fig-LbolcalcLEdd-vs-M} they are concentrated in the
  top-left region corresponding to small masses and
high bolometric luminosities. Further, NLS1s have the smallest $L_\nu^{\rm
disc}(5100\AA)/L_\nu^{\rm obs}(5100\AA)$ ratios compared to other objects
in the same range of BH   mass, as shown by Fig. \ref{fig-L5100calcobs-vs-M}.
  How to explain these
properties?

  We know that NLS1 galaxies have a strong soft X-ray excess compared to
   Broad Line Seyfert 1 galaxies (BLS1s) (Boller, Brandt \& Fink 1996 for a
review; see also the proceedings edited by Boller et al. 2000).
  This
would suggest that the BBB is shifted towards the EUV, and that the
ratio $L^{\rm disc}(5100\AA)/L^{\rm disc}({\rm EUV})$ is
smaller in NLS1s than in BLS1s. Meanwhile, this would mean that the ratio
$L_{\rm bol}/L^{\rm obs}(5100\AA)$ is larger in NLS1s than in BLS1s.
  However this is
impossible, as it would imply super-Eddington rates. Indeed
  Fig. \ref{fig-RoverRG-vs-L95100overLedd}, which
displays $R_{\rm BLR}/R_{\rm G}$ as a function of the Eddington ratio, for $L_{\rm
bol}/L^{\rm obs}(5100\AA)=9$, shows that with the conservative assumption of Eq.
\ref{eq:9}, NLS1s are already radiating at Eddington rates. NLS1s
  occupy a privileged position on this figure,
not only because of their small FWHM (corresponding by definition to
large $R_{\rm BLR}/R_{\rm G}$
values), but also because they display large Eddington ratios. Actually
this was strongly suspected on the basis of their spectral
and
variability properties in the X-ray range, similarly to galactic BH
candidates (Pounds, Done \& Osborne 1995).

{\it We conclude that NLS1s have not only larger Eddington
ratios, but also probably a larger ``non disc"
fraction of optical emission, compared to BLS1s}.

\section{What determines the size of the BLR ?}

\begin{figure}
\psfig{figure=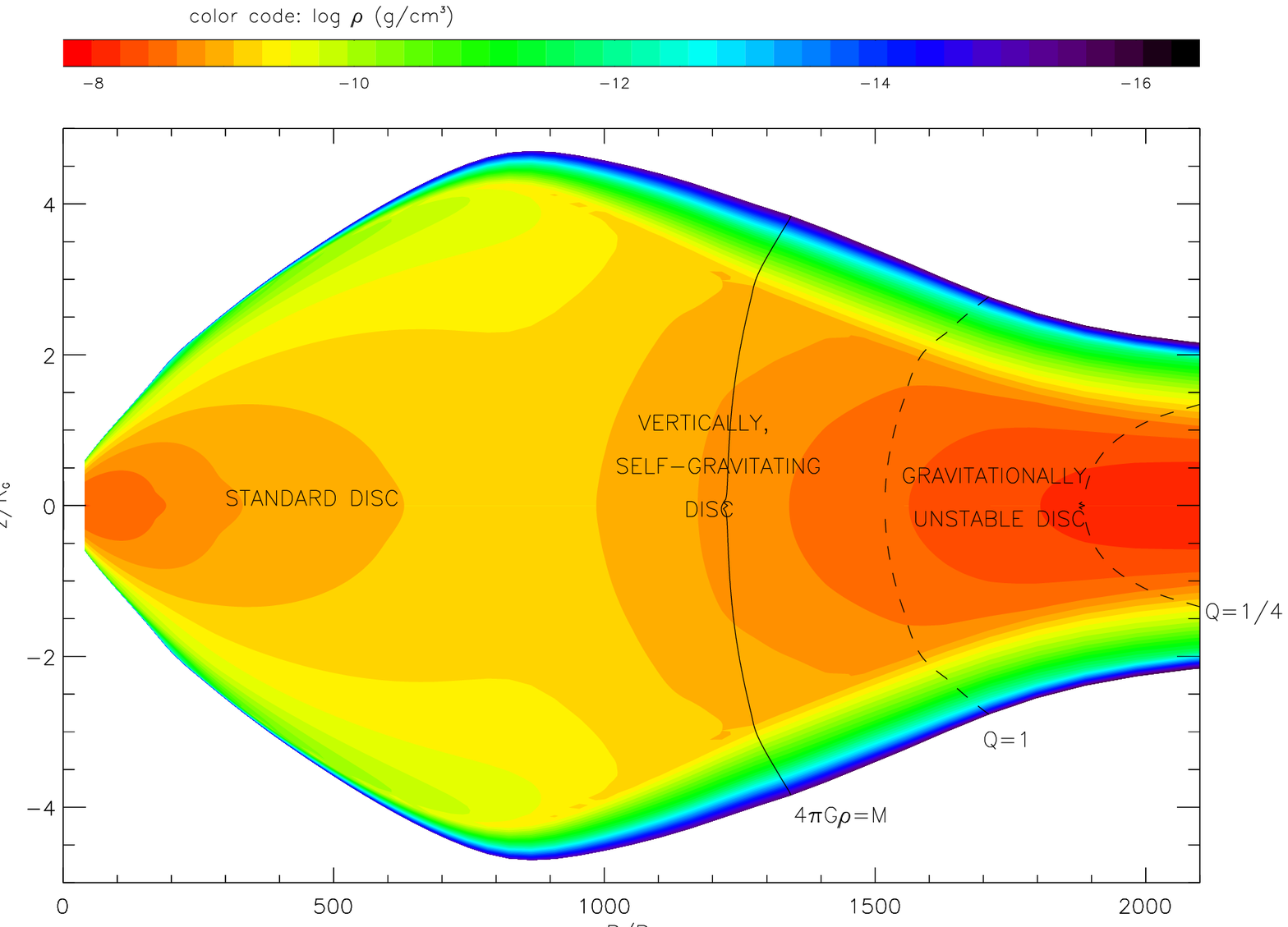,width=9cm,angle=0} \bigskip
\psfig{figure=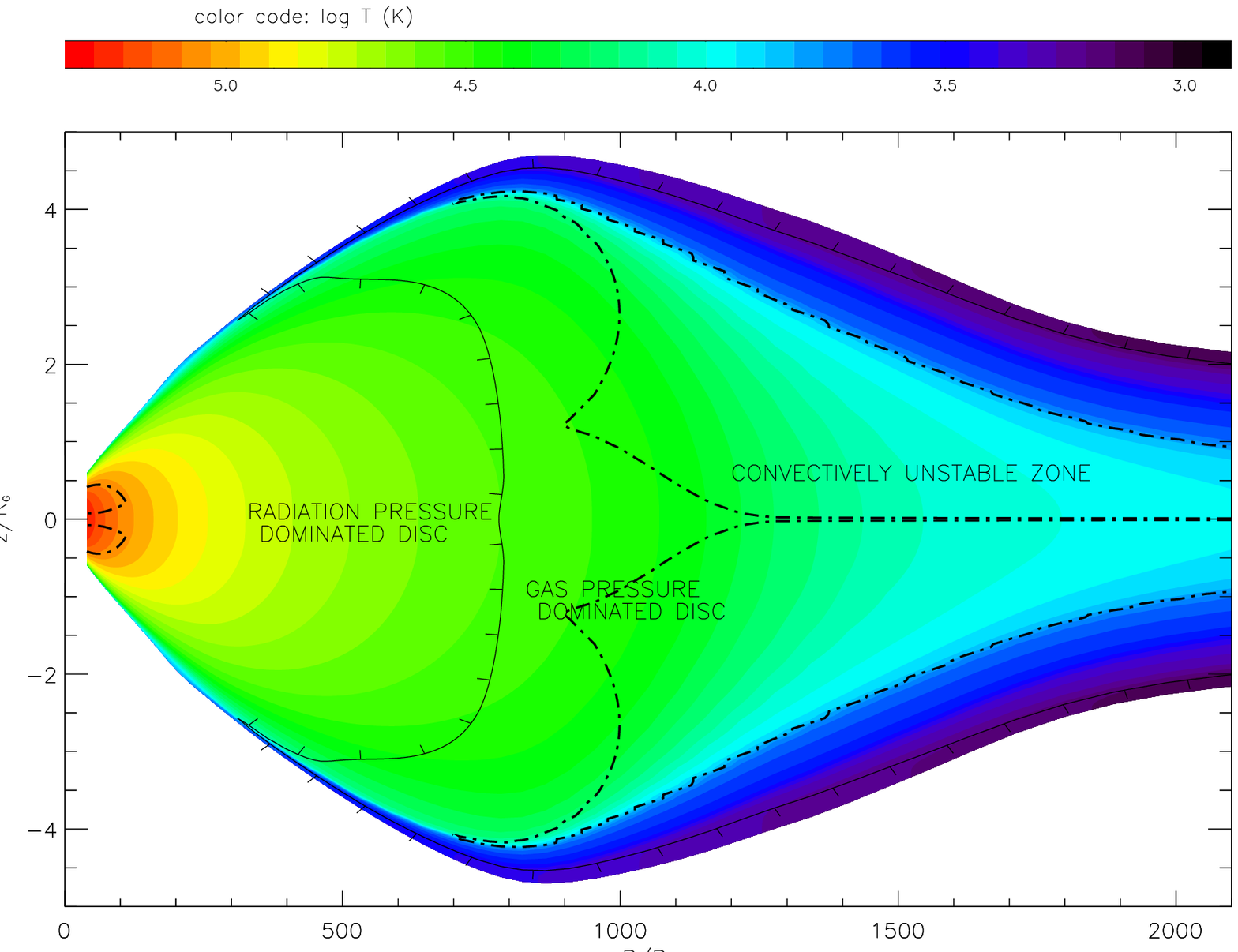,width=9cm,angle=0}
\caption{Density map ({\it top}) and temperature map ({\it bottom}) for a
steady state Keplerian $\alpha$-disc, surrounding a 10$^8$ M$_\odot$ black
     hole, versus $R$ and $z$ both normalized to $R_{\rm G}$. The accretion
rate is $\dot{M}=10^{-1}$ M$_\odot$/yr (i.e. $\dot{m} \sim
0.03$), and $\alpha=0.1$.}
\label{fig-2Dmap}
\end{figure}

\subsection{Model for the outer disc}

 Based mainly on distance arguments, we have suggested before that the BLR is
possibly related to the suppression of the disc by gravitational
instabilities. We have investigated this idea more quantitatively by a
series of bi-dimensional simulations of steady state, Keplerian accretion
$\alpha$-disc (Hur\'e 2000). More precisely, the model includes
convection in the framework of the Mixing Length Theory, turbulent
pressure that substantially thickens the disc and self-gravity within the
infinite disc approximation. Opacities and equation of state are realistic
and corresponds to a gas at LTE with cosmic abundances. External
irradiation is not taken into account, for several reasons. First the discs
considered here are optically very thick and moreover the irradiation flux
is likely to decrease as the gravitational
flux, having thus a relatively small influence on the disc structure
(Collin \& Hur\'e 1999). Second, as we shall see below, outer regions are in
the shadow of the
central regions due to self-gravity which pinches the disc vertically.
Figure \ref{fig-2Dmap} shows an example of internal structure of a disc
computed for $M=10^8$ M$_\odot$, $\dot{M}=10^{-1}$ M$_\odot$/yr and
$\alpha=0.1$. We see that the disc becomes self-gravitating (in the sense
$4\pi G \rho R^3 \ge M$) from $\sim 1200 \, R_{\rm G}$ which results in
a significant
decrease of the disc thickness and rise of the density. Note however that
the disc flaring vanishes at much lower radius $\sim 900 \, R_{\rm G}$.

\subsection{Location of the gravitationally unstable region}

\begin{figure}
\psfig{figure=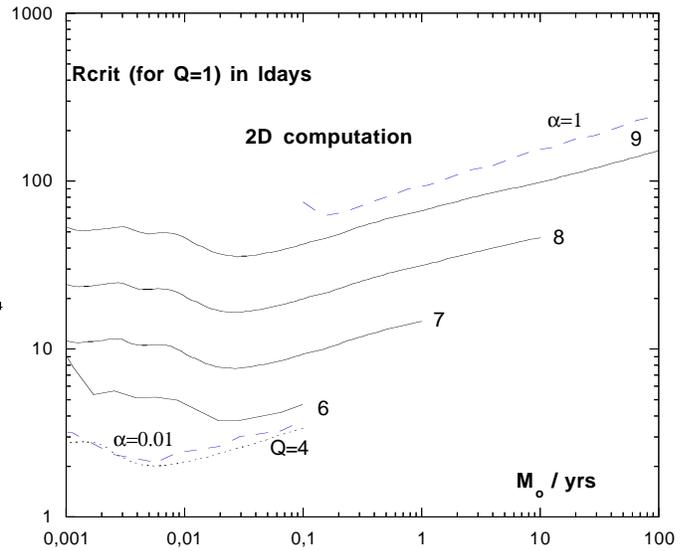,width=9cm,angle=0}
\caption{$R_{\rm crit}$ for $Q=1$ as a function of the
accretion rate expressed in M$_{\odot}$/yr. The curves are labeled with
the black hole mass in log(M$_\odot$). $\alpha=0.1$ throughout. A few other
cases are also displayed (for $\alpha=0.01$ and $\alpha=1$, and for
$Q=4$).}
\label{fig-Rcrit-vs-Mdot}
\end{figure}

In order to determine the radius $R_{\rm crit}$ where the disc becomes
gravitationally unstable, we have computed a series of disc models by
varying $M$ and $\dot{M}$. There are many possible sources of
uncertainties on the value of this quantity, including a lack of 
information regarding
some physical processes like turbulent viscosity through the
$\alpha$-prescription and its $z$-dependency, convective transport,
self-gravity, environment effects. In addition, the threshold value of the
$Q$-parameter which defines the instability criterion is uncertain in this
context (e.g., it is not clear whether this criterion applies to the
midplane quantities). We have attempted to estimate an absolute error on
$R_{\rm crit}$ by considering $0.01 \le \alpha \le 1$ and $ \frac{1}{4}
\le Q \le 4$ (Toomre, 1964; Goldreich \& Lynden-Bell 1965). Figure
\ref{fig-2Dmap} shows the location of the unstable disc region ($R_{\rm
crit} \sim 1500-2000 \, R_{\rm G}$). Our results are displayed in Figure
\ref{fig-Rcrit-vs-Mdot} which gives $R_{\rm crit}$ versus the
accretion rate for four mass decades, obtained for $\alpha=0.1$ and $Q=1$
(mean values). We see that $R_{\rm crit}$ is mainly determined by the
central mass and weakly sensitive to the accretion rate $\dot{M}$. The
$\alpha$-parameter and threshold value for $Q$ have rather minor effect
for moderate and high Eddington luminosities. The magnitude of the error
on $R_{\rm crit}$ is given by the gap between curves corresponding to the
same central mass but to different values of $\alpha$ and threshold values
for $Q$.
It is comforting to see that the results are quite reliable, as the
uncertainties do not produce an ``error'' larger than a factor say, $\sim
3$ on $R_{\rm crit}$.

\begin{figure}
\psfig{figure=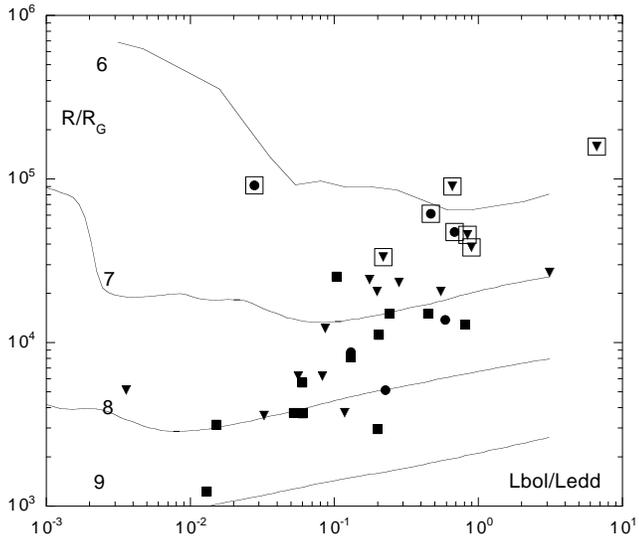,width=8.5cm,angle=0}
\caption{The points give $R_{\rm BLR}/R_{\rm G}$  versus $L_{\rm bol}/L_{\rm Edd}$, binned according to the
BH mass, and  assuming $L_{\rm bol}=9\times L(5100\AA)$, for the Kaspi et al. (2000) sample.
circles: $M\le 10^7$ M$_{\odot}$; triangles:  $10^7 \le M\le 10^8$ M$_{\odot}$;
squares:  $M\ge 10^8$ M$_{\odot}$.  The big open squares
are NLS1s.
The curves give $R_{\rm
crit}/R_{\rm G}$ from the 2D
simulations. They are labeled with the black hole mass in
log(M$_\odot$). }
\label{fig-RoverRG-vs-L95100overLedd}
\end{figure}

Figure \ref{fig-RoverRG-vs-L95100overLedd}
displays $R_{\rm crit}/R_{\rm G}$ versus the Eddington ratio for the same
models. A crude fit in the domain $0.01 \le \dot{m} \le 1$ gives
\begin{equation}
\frac{R_{\rm crit}}{R_{\rm G}} \approx  2\times 10^4 \, M_7^{-0.46}
\label{eq-Rsgbis}
\end{equation}

\subsection{Comparing the BLR and the disc sizes}

We have plotted on Fig. \ref{fig-R-vs-L5100} $R_{\rm crit}$ in light-days
versus $L(5100\AA)$, using Eq. (\ref{eq:9}) to convert
the disc accretion rate into optical luminosity (otherwise, accretion rates are
  unrealistically high),
together with the data
of the Kaspi et al. sample. Figure \ref{fig-RoverRG-vs-L95100overLedd}
displays $R_{\rm BLR}/R_{\rm G}$, for the Kaspi et al. sample,
still assuming Eq. (\ref{eq:9}). We see clearly from these
two figures that, in absolute, the BLR size is correlated with the
occurrence of the disc gravitational instability. So, if we identify the
edge of the gravitationally unstable disc with the inner edge BLR, that is
$R_{\rm crit} \le R_{\rm BLR}$, we conclude that {\it active nuclei
hosting low mass black holes are predicted to have broad lines narrower
than others}.

Fig. \ref{fig-RoverRG-vs-L95100overLedd} shows that, though the observed
masses do not match exactly the theoretical expectations,
there is a clear trend that objects with small masses (10$^6 < M < 10^7$
M$_{\odot}$) have a large $R_{\rm BLR}/R_{\rm crit}$ ratio and that objects
with large
masses ($M > 10^8$ M$_{\odot}$) have a small $R_{\rm BLR}/R_{\rm crit}$ ratio,
{\it in agreement with our
theoretical predictions}. So, at least to this ``zero order
approximation'', the gravitationally unstable region of the disc
  has kinematic and size properties compatible with the BLR
and could therefore be the ultimate source of the BLR clouds. It is however
difficult to draw more quantitative conclusions from this investigation.

\subsection{Influence of the ionization parameter of the BLR}

One can notice that $R_{\rm crit}$ depends mainly of the BH   mass and
little on the luminosity, while a correlation appears on Fig.
\ref{fig-R-vs-L5100}
between $R_{\rm BLR}$ and the $L^{\rm obs}(5100\AA)$. Kaspi et al. (2000)
represent this correlation as $R_{\rm BLR} \propto
L(5100\AA)^{0.70}$. A smaller power law index (closer to $0.5$)
was found in previous works
(e.g. Wandel et al. 1999). Anyway it is not clear how to translate such a
relation into a relation between $R_{\rm BLR}$ and $L_{\rm bol}$, because
of an important uncertainty arising from the optical-to-bolometric
conversion. If the optical to bolometric luminosity ratio is constant, it
results in $R_{\rm BLR} \propto L_{\rm bol}^{0.5-0.7}$.

  In a recent paper, Nicastro (2000) proposed that the BLR clouds are 
released by the
accretion disc in the region where a vertically outflowing
corona exists, according to a model proposed by Witt, Czerny \& Zicky
(1997). This is indeed an interesting model, since in this case the 
main parameter
governing the size of the BLR would be $\dot{m}$ and not $M$. However one
would expect that the remote regions giving
rise to the BLR, which are strongly pressurized by the corona, are also
gravitationally unstable.
Moreover there is also another important parameter governing the size 
of the BLR,
that is its ionization parameter
\begin{equation}
\xi=\frac{L_{\rm ion}}{n R_{\rm BLR}^2}
\end{equation}
where $L_{\rm ion}$ is the ionizing luminosity ($\sim L_{\rm bol}$) and
$n$ is the number density in the clouds. Actually, any model must
reproduce not only the kinematics and dimension of the BLR but also the
ionization parameter (linked to the ionization state) consistent with the
observed line ratios which are almost independent on the luminosity over
more than four
decades (except for the Baldwin effect which might
indicate a decrease of the ionization parameter with increasing
luminosity, but only for high ionization lines).
Typically, $\xi \approx 1$ and detailed studies of the line ratios show
that $n \approx 10^{10}$ cm$^{-3}$. If the density is about a constant,
the constancy of $\xi$
translates into a relation  $R_{\rm BLR} \propto L_{\rm bol}^{0.5}$, in
close agreement with the observations.
Another way to express  this relation is
that the flux irradiating the clouds is constant.
This relation has been used in the ``ionization method" aiming to
determine the mass of the black holes, and it gives results compatible with
the reverberation
mapping method (Wandel, Peterson \& Malkan 1999), but more uncertain.

Thus  a possible interpretation of the observed correlation
is that among clouds released by the gravitationally unstable disc, only
those
   located at the right distance (i.e. corresponding to the right
ionization parameter or ionizing flux) are observed as BLR clouds: the
others could contribute to the Warm Absorber. This
could explain why $R_{\rm crit}$ constitutes a ``lower envelope" to $R_{\rm BLR}$
  for a given BH mass. Also since a constant
ionizing flux corresponds to $R_{\rm BLR}/R_{\rm G}\propto
\sqrt{\dot{m}/M}$, we
see that the objects displaying the largest $R_{\rm BLR}/R_{\rm G}$ 
ratios, hence
the smaller line widths, are those which have the largest accretion rates
in Eddington unit and the smallest masses. Such a situation is proposed to 
explain the properties of NLS1s.
Note that in this picture, the more intense soft X-ray excess 
displayed by NLS1s compared
to BLS1s can also play a role, since such a spectrum has a stronger ionizing
power for the same ionization parameter
(e.g. Wandel 1997).

\section{Conclusion}

        We have discussed the relations between the optical
luminosity, the BH mass, and the size of the BLR, in the framework of the
standard accretion disc model.

We have shown first that the
optical luminosity cannot be accounted for, meaning that the standard
accretion disc picture (stationary, geometrically thin and optically
thick) does not hold, at least in the region emitting the optical band,
i.e. at $\sim 100R_{\rm G}$.  Either the major
  fraction of
the optical luminosity is not due to disc emission, or the disc is
not ``standard": it is
unstable, or the accretion rate depends on the radius, owing to strong
outflows or to advection dominated accretion. In these last cases, the
implied mass rate would have to be strongly super-critical. We would thus favor
  the first explanation. We have also
shown that NLS1s are extreme in this context:  either they have a larger
fraction of non disc emission in the optical range and are radiating close
to their Eddington luminosity, or they have strongly super-Eddington
luminosity, which seems implausible. So again in this case it seems that
the best explanation would be that a large fraction of the optical
emission (larger even than for BLS1s) is not produced by the disc.

In the second part, we have studied the relation between the disc and the
BLR, and we have shown that there is a good agreement between the size of
the BLR and the critical radius at which an $\alpha$-disc becomes
gravitationally unstable. This suggests that the BLR is produced
above the gravitationally unstable part of the disc. However the ionization
parameter should also play a role in explaining the correlation observed
between the size and the luminosity. If BLR clouds are seen only for a
small range of ionization parameter or of ionizing flux, the size of the
BLR expressed in $R_{\rm G}$ should increase with the accretion rate 
expressed in
Eddington unit and decrease with the BH mass, according to the observed
correlation. This could explain the small widths of NLS1s.

\begin{acknowledgements}

We are grateful to Martin Ward for a careful reading of the manuscript 
leading to several improvments.

\end{acknowledgements}

\end{document}